\documentclass[%
 reprint,
 amsmath,amssymb,
 aps,
]{revtex4-2}
\usepackage{float}
\usepackage{graphicx}
\usepackage{dcolumn}
\usepackage{bm}

\begin{document}

\title{Comparing the buckling strength of spherical
shells with dimpled versus bumpy defects}

\author{Arefeh Abbasi}
\affiliation{%
 Ecole Polytechnique Fédérale de Lausanne (EPFL)\\
    Flexible Structures Laboratory\\
    CH-1015 Lausanne, Switzerland
}
    
\author{Fani Derveni}%
\affiliation{%
 Ecole Polytechnique Fédérale de Lausanne (EPFL)\\
    Flexible Structures Laboratory\\
    CH-1015 Lausanne, Switzerland
}%

\author{Pedro M. Reis}%
 \email{pedro.reis@epfl.ch}
\affiliation{%
 Ecole Polytechnique Fédérale de Lausanne (EPFL)\\
    Flexible Structures Laboratory\\
    CH-1015 Lausanne, Switzerland
}%


\begin{abstract}
We investigate the effect of defect geometry in dictating the sensitivity of the critical buckling conditions of spherical shells under external pressure loading. Specifically, we perform a comparative study between shells containing dimpled (inward) versus bumpy (outward) Gaussian defects. The former has become the standard shape in many recent shell-buckling studies, whereas the latter has remained mostly unexplored. We employ finite-element simulations, which were validated previously against experiments, to compute the knockdown factors for the two cases while systematically exploring the parameter space of the defect geometry. For the same magnitudes of the amplitude and angular width of the defect, we find that shells containing bumpy defects consistently exhibit significantly higher knockdown factors than shells with the more classic dimpled defects. Furthermore, the relationship of the knockdown as a function of the amplitude and width of the defect is qualitatively different between the two cases, which also exhibit distinct post-buckling behavior. A speculative interpretation of the results is provided based on the qualitative differences in the mean-curvature profiles of the two cases.
\end{abstract}
\maketitle
\section{\label{sec:level1}Introduction}
\label{sec:introduction}
The mechanical response of thin elastic shells under compression is highly nonlinear~\cite{Niordson1985,Koiter1969}, with a strong sensitivity to imperfections~\cite{von1939buckling,von1940influence, koiter1945over, hutchinson1970postbuckling}. Predicting the buckling strength of shells is a longstanding canonical problem in the structural mechanics community~\cite{Elishakoff2014,samuelson2003shell}. The classic prediction for the perfect spherical shell case was first obtained by Zoelly in 1915~\cite{Zoelly1915} from linear buckling analysis:
\begin{equation}
    p_\mathrm{c}=\frac{2E}{\sqrt{3(1-\nu^2)}}\eta^{-2},
    \label{eq:zoelly}
\end{equation}
where $E$, $\nu$, and $\eta= R/h$ are the Young’s modulus, Poisson’s ratio, and slenderness ratio of the shell of radius $R$ and thickness $h$, respectively. However, experimental measurements for the critical buckling pressure of a thin spherical shell containing imperfections~\cite{seaman1962nature, kaplan1954nonlinear, tsien1942theory, krenzke1963elastic, babcock1983shell, Carlson1967,lee2016geometric} are always found to be lower than the theoretical prediction in Eq.~(\ref{eq:zoelly}) due to their extreme sensitivity to imperfections~\cite{koiter1945over}. The discrepancies between theory and experiments have been attributed to the non-uniformity of loading~\cite{bijlaard1960elastic}, the boundary conditions~\cite{kobayashi1966influence}, the influence of pre-buckling deformations~\cite{almroth1966influence}, and the deviations from perfect shell geometry~\cite{budiansky1972buckling}. The ratio between the measured critical pressure, $p_\mathrm{max}$, and the corresponding classic prediction for the perfect geometry, $p_\mathrm{c}$ is known as the 
\textit{knockdown factor},
\begin{equation}
 \kappa=\frac{p_\mathrm{max}}{p_\mathrm{c}}, 
 \label{knockdown}
\end{equation}
which is always smaller than unity ($\kappa<1$) for realistic shells that inevitably contain material and geometric imperfections.  
Despite the classic, albeit still challenging, nature of the problem, there has been a recent revival in the interest and research of shell buckling. The study of the critical buckling conditions of spherical shells has been reinvigorated by recent advances in experiments and computation~\cite{lee_fabrication_2016, lee2016geometric, marthelot_buckling_2017, yan2020buckling, lee2019evolution, virot_stability_2017, gerasimidis_establishing_2018, fan2019critical, Lazarus2012}. For a contemporary perspective and overview of the recent activity in the field, we point the reader to the following recent studies~\cite{lee2016geometric, jimenez_technical_2017, marthelot_buckling_2017, yan2020buckling, lee2019evolution, abbasi2021probing, pezzulla2019weak, derveni2022probabilistic, hutchinson_imperfections_2018,hutchinson2017nonlinear,hutchinson_john_w._nonlinear_2017, hutchinson2018imperfections}. Even if similar results are also found for cylindrical shells~\cite{Homewood1961,Elishakoff2014, von1941buckling, bijlaard1960elastic, budiansky1972buckling, virot_stability_2017, fan2019critical, groh2017exploring, groh2019spatial}, the present study will focus on spherical shells exclusively.

\begin{figure}[h!]
    \centering
    \includegraphics[width=0.9\columnwidth]{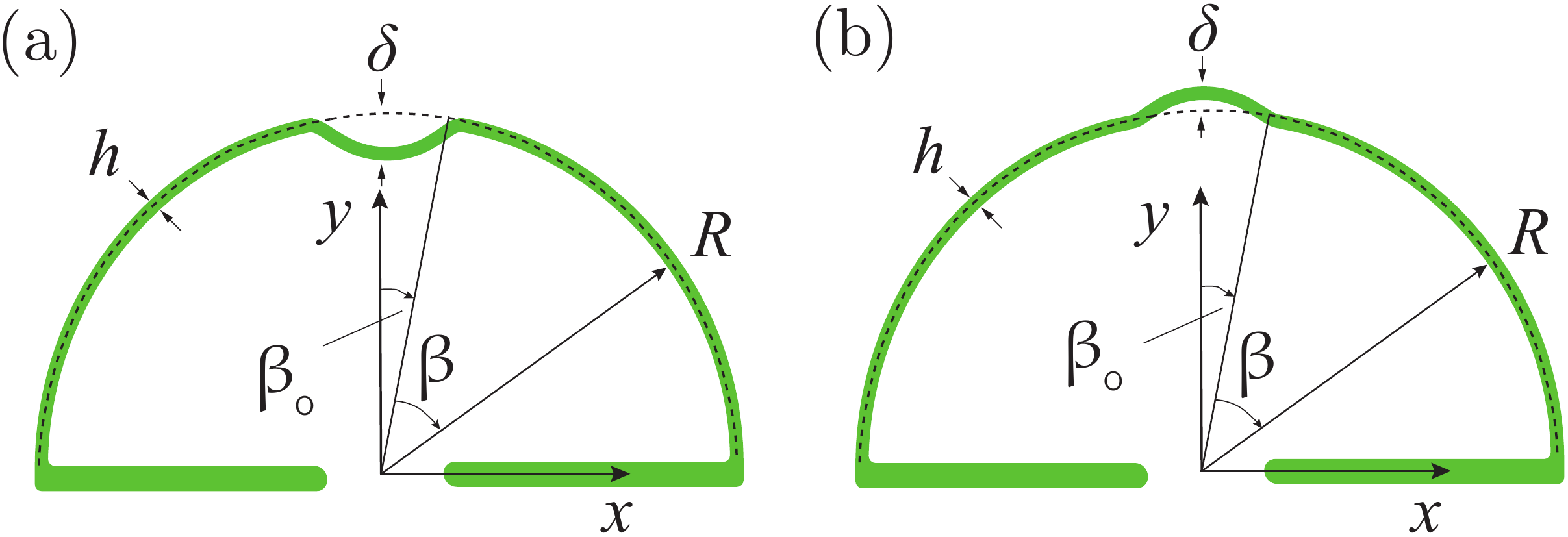}
    \caption{
    Schematic diagrams of the two types of geometry considered for our imperfect shells containing (a) a dimpled defect, and (b) a bumpy defect, with $c=-1$ and $c=+1$ (cf. Eq.~\ref{eqn:gaussiandimple}), respectively. In both cases, the hemispherical shells have radius $R$ and thickness $h$, and the defect is located at the pole ($\beta_{\circ}$) with a geometry characterized by the amplitude, $\delta$, and half-angular width, $\beta_\circ$.
    } 
    \label{fig:fig1}
\end{figure}

Most of the recent investigations on spherical-shell buckling mentioned in the previous paragraph~\cite{hutchinson2016buckling, hutchinson_john_w._nonlinear_2017, hutchinson2017nonlinear, hutchinson_imperfections_2018, thompson2016shock,thompson2017probing} have considered standardized dimpled (Gaussian) defects. Other types of imperfections (\textit{e.g.}, through-thickness defects~\cite{yan2020buckling,hutchinson_effect_1971,paulose2013buckling}, and dent imperfections~\cite{gerasimidis2021dent}) have also been considered, but such cases are sparser. A benefit of focusing on standardized dimples is that they allow for a better contextualization and interpretation of results across different studies. These dimpled imperfections are axisymmetric, localized, and characterized by a radial modulation of the shell mid-surface from a perfect sphere of radius $R$, by  
\begin{equation} \label{eqn:gaussiandimple}
w_I=c \delta e^{-\left(\beta / \beta_\circ \right)^2}, 
\end{equation}
where $\beta$ is the polar angle measured from the north pole ($\beta_{\circ}$ where the center of the defect is located), and the constants $\beta_{\circ}$ and $\delta$ control the width and amplitude of the defect (see Fig.~\ref{fig:fig1}). The defect amplitude, which is typically normalized by the thickness of the shell, $\overline{\delta}=\delta/h$, corresponds to the maximum radial deviation at the center of the defect. It is also common to define a geometric parameter ~\cite{koga1969axisymmetric},
\begin{equation}
\lambda=\left\{12\left(1-\nu^2\right)\right\}^{1 / 4} \eta^{1 / 2} \beta_\circ,
\label{eq:lambda}
\end{equation} 
to rescale the defect width, normalizing effects arising from the radius-to-thickness ratio, $\eta$, of the shell.

In the existing literature, the prefactor $c$ in Eq.~(\ref{eqn:gaussiandimple}) has been consistently set to $c=-1$, corresponding to inward-pointing dimples, as shown schematically in Fig.~\ref{fig:fig1}(a). The knockdown factor, $\kappa$, of such shells containing dimpled defects was found in experiments, as well as theoretical and computational analyses, to depend strongly on $\overline{\delta}$, dropping sharply from unity for $0 < \overline{\delta} \lesssim 1$ and reaching a plateau for $\overline{\delta} \gtrsim 1$~\cite{lee2016geometric}.  Moreover, these results demonstrated that the geometric parameter $\lambda$ governs the onset and level of the plateau in the $\kappa(\overline{\delta})$ curves, as characterized thoroughly in Ref.~\cite{jimenez_technical_2017}. The authors revealed a lower
bound of the plateau level that depends solely on $\eta$ and $\lambda$.

Here, we revisit the buckling of a spherical shell containing a single Gaussian defect according to Eq.~(\ref{eqn:gaussiandimple}). We perform a comparative study of the knockdown factor for the previously considered dimpled (inward) defects ($c=-1$; see Fig.~\ref{fig:fig1}a) compared to the symmetric case for bumpy (outward) defects ($c=+1$; see Fig.~\ref{fig:fig1}b). Recently, Derveni \textit{et al.}~\cite{derveni2022probabilistic} have studied the buckling of shells containing a large distribution of defects, validating FEM simulations against experiments using bumpy defects, a choice that was driven by practical experimental constraints; but the difference between dimples and bumps was not explored in detail. Otherwise, to the best of our knowledge, bumpy defects have not been investigated systematically to date. We will focus on the following research question: How does the buckling strength compare between single-imperfection shells containing a dimpled versus a bumpy defect?
\section{\label{sec:level1}Methodology: Finite element analysis}
\label{sec:methods}
\begin{figure*}[ht!]
    \centering
    \includegraphics[width=1\textwidth]{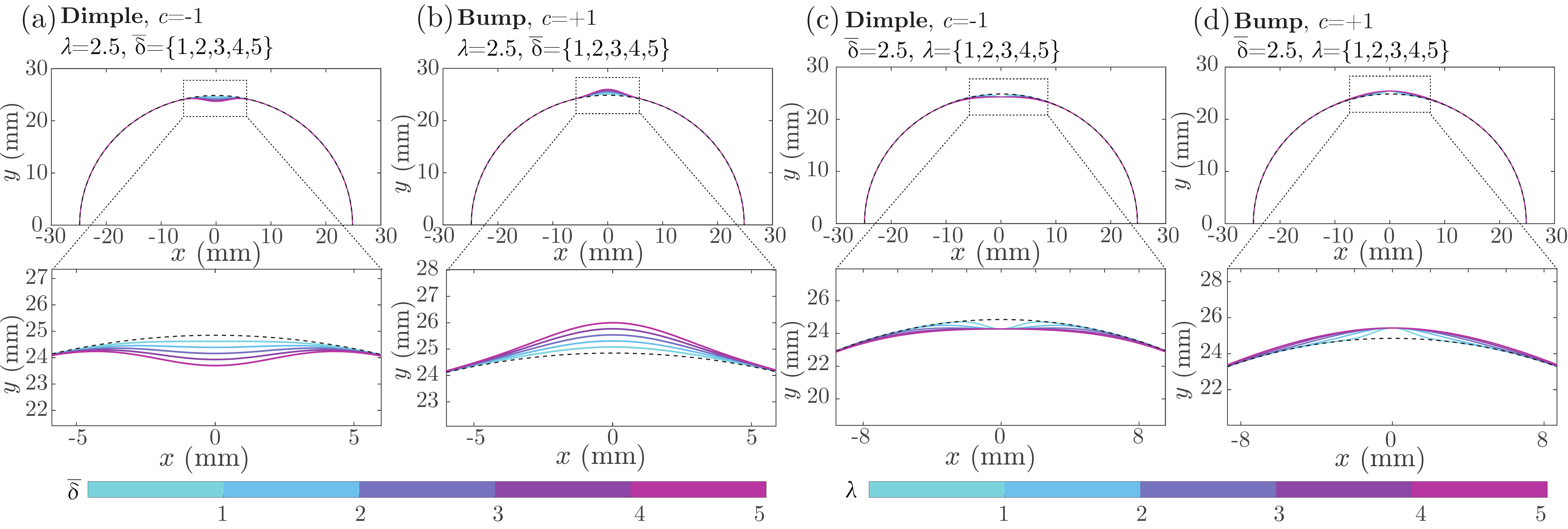}
    \caption{Representative examples of the initial geometric profiles of the imperfect shells considered. The shells contain defects with (a,b) $\lambda=2.5$ and $0 \leq \overline{\delta} \leq 5$, and (c,d) $\overline{\delta}=2.5$ and $0 \leq \lambda \leq 5$. The defects correspond to $c=-1$ in (a,c) and $c=+1$ in (b,d). The lower panels show amplified views near the defects. These geometric profiles serve as input to the FEM simulations.
    } 
    \label{fig:fig2}
\end{figure*}

In Fig.~\ref{fig:fig1}, we present schematic diagrams of the two types of geometries for the imperfect hemispherical shells that we will consider, containing either a dimpled defect ($c=-1$ in panel (a)) or a bumpy defect ($c=+1$ in panel (b)). We will focus on hemispherical shells of radius $R=24.85\,$mm, thickness $h=0.23\,$mm, and, thus, $\eta= R/h = 108$ with a single imperfection located at the pole, without loss of generality~\cite{hutchinson2016buckling} given the large value of $\eta$. This generality assumes there is essentially no dependence of knockdown factor characterization on $\eta$ for sufficiently slender shells as long as the defect width is scaled according to Eq.~(\ref{eq:lambda}). Each shell is clamped at the equator and (de)pressurized to load it under compression until buckling occurs. 

The initial shell geometry considered in the simulations is axisymmetric. As such, the 2D cross-sectional profiles of the imperfect shells presented in Fig.~\ref{fig:fig2} for different values of $\overline{\delta}$ and $\lambda$ (see color bar) suffice to fully describe this initial geometry. The perfectly spherical case ($\overline{\delta}=0$, $\lambda=0$) is represented by the dashed line. Panels (a,\, c) and (b,\, d) represent the shell with dimpled ($c=-1$) and bumpy ($c=+1$) defects, respectively. Representative defects with the same defect width, $\lambda=2.5$, in a range of amplitudes, $\overline{\delta} \in \{1, 2, 3, 4, 5\}$, are shown in Figs.~\ref{fig:fig2}(a,b). In Figs.~\ref{fig:fig2}(c,d), we present representative shell profiles with the same defect amplitude, $\overline{\delta}=2.5$, in a range of widths, $\lambda \in \{1, 2, 3, 4, 5\}$. The corresponding lower panels in Figs.~\ref{fig:fig2} show  magnified views of the defect profiles localized at the pole. Beyond these representative cases, our investigation will consider the following ranges for the geometric-parameters space of the defect:
$\overline{\delta}\in[0.1,\,5]$ in steps of $\Delta \overline{\delta}=0.1$ for the defect amplitude and $\lambda\in[0.25,\,5]$ in steps of $\Delta \lambda=0.25$ for $\lambda \leq 1$ and $\Delta \lambda=0.5$ for $\lambda \geq 1$ for the (normalized) defect width; while fixing all other parameters mentioned above. Although these initial geometries are  axisymmetric, it is important to anticipate, as our results will evidence, that the post-buckling modes can be asymmetric, especially for shells with bumpy defects.

The material was modeled as a neo-Hookean solid, with Young’s modulus of $E=1.26\,$MPa, and a Poisson's ratio of $\nu \approx 0.5$ (assuming incompressibility). These material-specific material properties were chosen to align with the previous experimental studies in Refs.~\cite{lee2016geometric, marthelot_buckling_2017, yan2020buckling, abbasi2021probing, derveni2022probabilistic}, where they were measured directly from experiments and used to validate the finite-element simulations.

The set of geometric and physical parameters mentioned above was chosen to match with Ref.~\cite{lee2016geometric} toward enabling a direct comparison with this previous study. However, for the present simulation framework, instead of using the axisymmetry model of Refs.~\cite{lee2016geometric, jimenez_technical_2017}, we use a three-dimensional description of the structure using shell elements to capture possible asymmetry buckling behavior. This finite element modeling (FEM) approach has been validated against precision experiments for the specific problem of shell buckling~\cite{derveni2022probabilistic,abbasi2021probing}. We followed the same FEM methodology to perform simulations with the commercial package Abaqus/Standard; the details are given in Ref.~\cite{derveni2022probabilistic}. We employed four-node S4R shell elements with reduced integration points to discretize the shell using sweep meshing, with 300 and 1200 elements in meridional and azimuthal directions, respectively. A mesh convergence study was also conducted in order to ensure that the results were not influenced by mesh size. A Riks solver~\cite{riks1979incremental} was used to capture the progress of the simulation along the arc length of the load-displacement curve. Geometric nonlinearities were considered throughout the study.

In the FEM simulations, each imperfect shell geometry was pressurized until the onset of buckling, at which point the maximum pressure value, $p_\mathrm{max}$, was recorded. Then, the knockdown factor was computed using Eq.~(\ref{knockdown}). Throughout the manuscript, for ease of comparison and to confusion, we will refer to the knockdown factor of the imperfect shell with a dimpled defect as $\kappa_\mathrm{D}$ and $\kappa_\mathrm{B}$ for the bumpy defect. The FEM results for the dimpled shells were first verified against Ref.~\cite{lee2016geometric} in the previously explored range of parameters and then expanded to a systematic parameter exploration of dimpled and bumpy defects.
\section{\label{sec:level1}Results}
\label{sec:results}
\begin{figure}[h!]
    \centering
    \includegraphics[width=0.95\columnwidth]{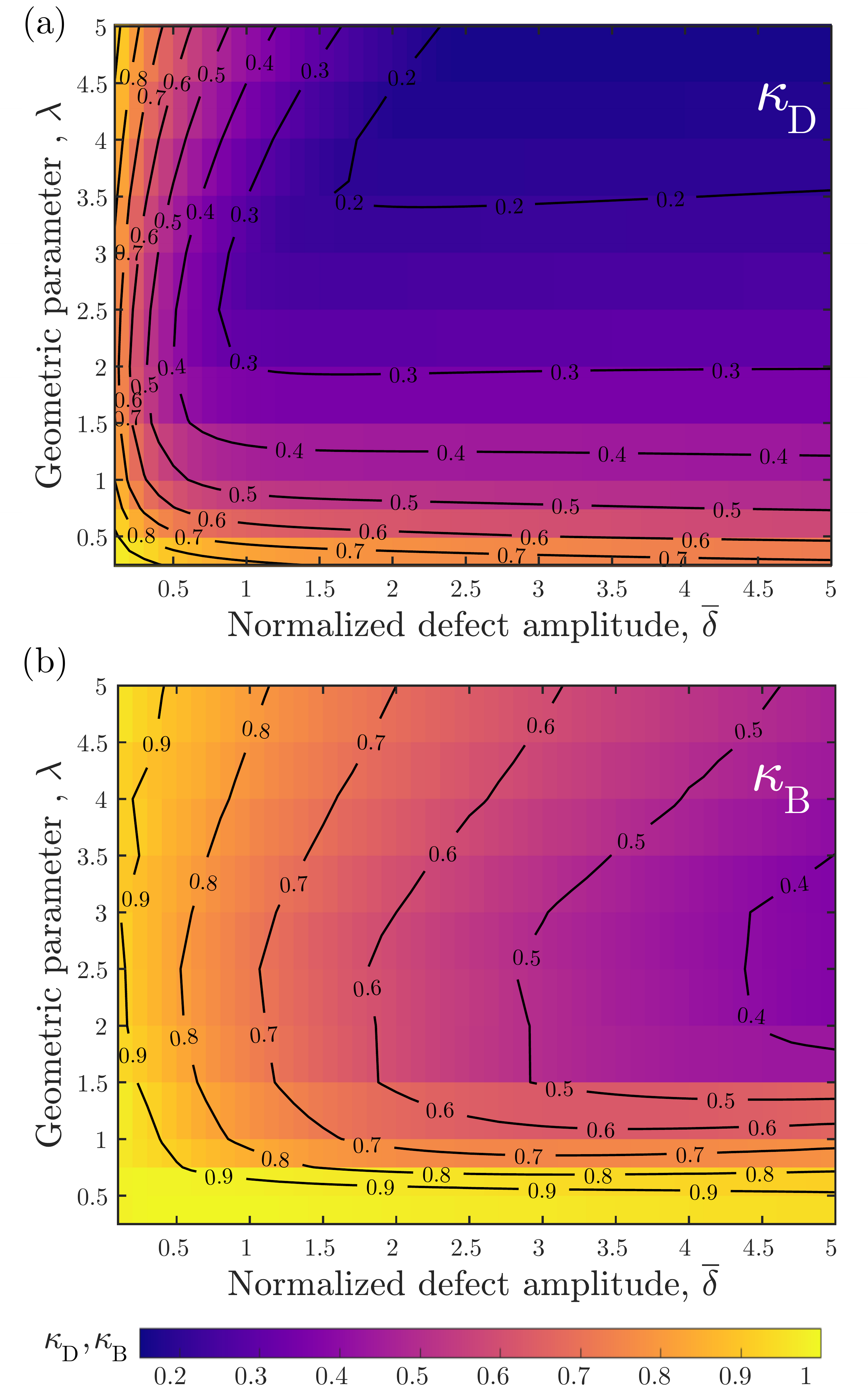}
    \caption{Surface plots of the knockdown factor of shells containing (a) a dimpled imperfection, $\kappa_\mathrm{D}$, and (b) a bumpy imperfection, $\kappa_\mathrm{B}$, for different values of the dimensionless geometric parameter (width), $0.25\leq \lambda \leq 5$, and normalized defect amplitude, $0.1 \leq \overline{\delta} \leq 5$. Counter lines are superposed for the corresponding values of $\kappa_\mathrm{D}$ and  $\kappa_\mathrm{B}$, in steps of 0.1. The color bar is shared for (a) and (b).}
    \label{fig:fig_3}
\end{figure}

Following the methodology introduced above, we start our investigation to explore the parameter space of dimpled and bumpy defects. We will characterize and compare the effects of bumps and dimples on the buckling behavior, especially the knockdown factor, of the pressurized imperfect shells.    

In Fig.~\ref{fig:fig_3}, we present surface plots with all the data we obtained from the FEM simulations for the knockdown factor of shells with a dimpled and bumpy imperfection in the whole parameter space $(\overline{\delta},\, \lambda)$ specified in Section~\ref{sec:methods}: panel (a) for  $\kappa_\mathrm{D}$ and panel (b) for $\kappa_\mathrm{B}$. Color coding is used to quantify the knockdown factor (see the colorbar). Contour lines for constant values of $\kappa_\mathrm{D}$ and $\kappa_\mathrm{B}$, in intervals of 0.1, are superposed on the surface plots. For the dimpled shells (Fig.~\ref{fig:fig_3}a), the minimum value of the knockdown factor, $\kappa_\mathrm{D} \approx 0.15$, is found on the upper extremity of the $(\overline{\delta},\, \lambda)$ parameter space. This means that a shell with the deepest and widest defect has the lowest knockdown factor, a fact that is well-established in the literature. By contrast, for bumpy shells (Fig.~\ref{fig:fig_3}b), in the explored range, the minimum knockdown factor ($\kappa_\mathrm{B} \approx 0.37$) occurs for the defects with the largest amplitude but intermediate width ($2\lesssim \lambda \lesssim 3$). Overall, the values of $\kappa_\mathrm{B}$ are consistently larger than those of $\kappa_\mathrm{D}$; the geometry of dimples plays a more significant role in reducing the knockdown factor of an imperfect shell compared to bumps. These features highlight the first and major qualitative differences between the two cases. 
\begin{figure}[h!]
    \centering
    \includegraphics[width=0.9\columnwidth]{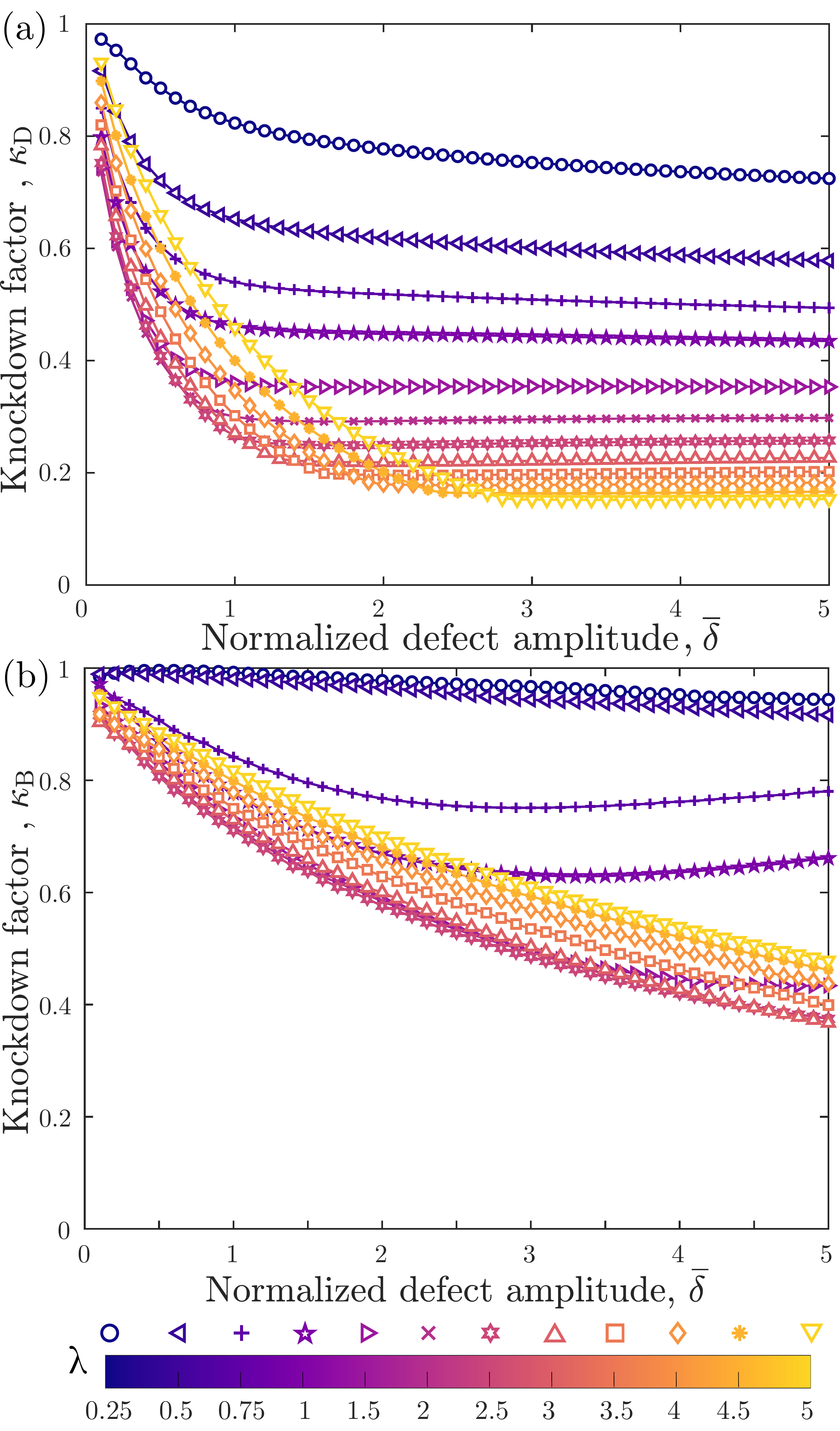}
    \caption{Knockdown factor, $\kappa$, as a function of the normalized defect amplitude, $\overline{\delta}$, for imperfect shells with defects in a range of $\lambda\in[0.25,\,5]$ (see colorbar and marker symbols). 
    (a) Knockdown factor, $\kappa_\mathrm{D}(\overline{\delta})$, for a shell with a dimpled imperfection; \textit{i.e.}, $c=-1$  in Eq.~(\ref{eqn:gaussiandimple}). 
    (b) Knockdown factor, $\kappa_\mathrm{B}(\overline{\delta})$, for a shell with a bumpy imperfection; \textit{i.e.}, $c=+1$ in Eq.~(\ref{eqn:gaussiandimple}).  
    }
    \label{fig:fig_4}
\end{figure}

Next, we elaborate on the data presented in Fig.~\ref{fig:fig_3} to more comprehensively describe the impact of the various parameters of the defect geometric on the knockdown factor, $\kappa_\mathrm{D}$ for dimples and $\kappa_\mathrm{B}$ for bumps. For this purpose, we first characterize the dependence of the knockdown factor on the defect amplitude and then on the normalized defect width, for both cases. 

In Fig.~\ref{fig:fig_4}, we present $\kappa_\mathrm{D}$ and $\kappa_\mathrm{B}$ as functions of $\overline{\delta}$, each curve corresponding to a different value of $\lambda$ (see colorbar and marker symbols). The data for shells with dimpled imperfections is shown in panel (a), and those with bumpy imperfections in panel (b). 

Note that the $\kappa_\mathrm{D}(\overline{\delta})$ data in Fig.~\ref{fig:fig_4}(a) are a recomputation of what is already presented in Ref.~\cite{lee2016geometric}, while the range of geometric parameters for $\lambda<1$ and $\overline{\delta}>3$ is further expanded herein. Still, for verification purposes, we selected a  specific set of parameters ($ \lambda(0<\overline{\delta}\leq 3)=\{1.5,\, 5\}$) and confirmed identical results to those in Ref.~\cite{lee2016geometric}. We recall that in this previously studied case of dimpled imperfections, $\kappa_\mathrm{D}$ decreases monotonically with $\overline{\delta}$ and eventually reaches a plateau. Both the plateau level and its onset depend on $\lambda$, as characterized previously in Ref.~\cite{jimenez_technical_2017}. The plateau is less pronounced when $\lambda < 1$ (regime not explored previously). For example, in the extreme case of $\lambda =0.25$ (the narrowest defects), no plateau is reached; after an initially fast decay, the knockdown factor continues to decrease as the amplitude increases all the way to high-amplitude defects of $\overline{\delta}=5$. We emphasize that there is little novelty in these results for dimpled shells, which were already presented in Ref.~\cite{lee2016geometric} and are presented here for completeness to enable a direct comparison with the case of bumpy imperfections discussed next.

Imperfect shells with bumpy defects exhibit a $\kappa_\mathrm{B}(\overline{\delta})$ behavior (Fig.~\ref{fig:fig_4}b) that is qualitatively different from the dimpled case discussed above (Fig.~\ref{fig:fig_4}a). The main  feature is that the values of $\kappa_\mathrm{B}$ tend to be  higher overall than $\kappa_\mathrm{D}$, with smoother decays as a function of $\overline{\delta}$, and non-monotonic behavior in some of the curves. Moreover, the $\kappa_\mathrm{B}$ curves do not exhibit the prominent plateaux observed in $\kappa_\mathrm{D}$. Three regimes are observed. First, for shells with relatively narrow defects, $\lambda=\{0.25, 0.5\}$, $\kappa_\mathrm{B}$ remains close to unity across the entire range of $\overline{\delta}$; these shells are nearly insensitive to imperfections. Second, for shells with intermediate-width defects, $\lambda=\{0.75, 1\}$, the $\kappa_\mathrm{D}(\overline{\delta})$ curves are non-monotonic; $\kappa_\mathrm{B}$ decreases for $0.1 \leq \delta \lesssim 3$ and then increases beyond $\overline{\delta}\approx 3$. Third, for $\lambda \geq 1.5$, $\kappa_\mathrm{B}(\overline{\delta})$ decreases again monotonically. 

\begin{figure}[b!]
    \centering
    \includegraphics[width=0.9\columnwidth]{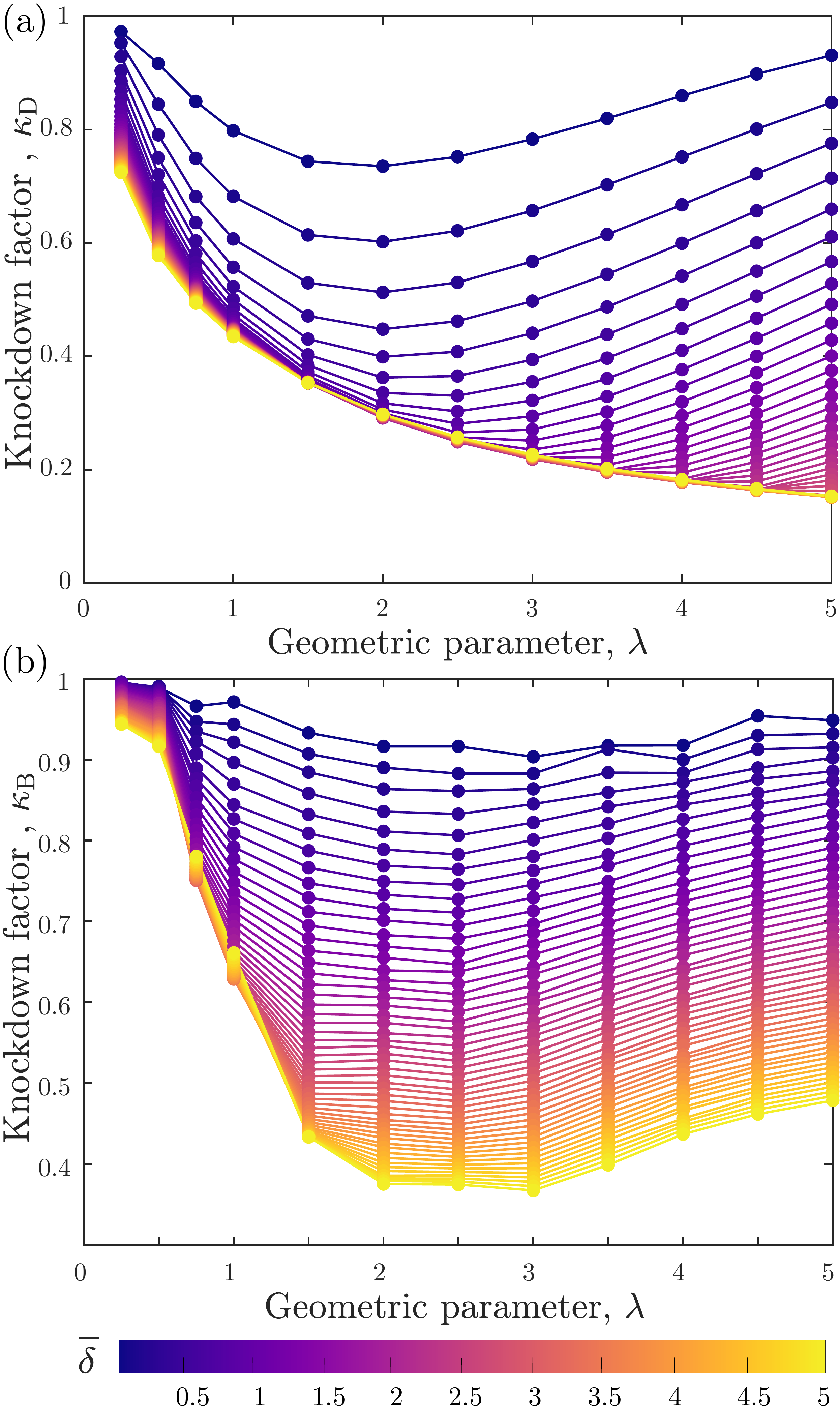}
    \caption{Knockdown factor, $\kappa$, as a function of the normalized defect width, $\lambda$ for imperfect shells with defect amplitudes in a range of $\overline{\delta}\in[0.1,\,5]$ (see colorbar). (a) Knockdown factor, $\kappa_\mathrm{D}(\lambda)$, for a shell with a dimpled imperfection; \textit{i.e.}, $c=-1$  in Eq.~(\ref{eqn:gaussiandimple}). (b) Knockdown factor, $\kappa_\mathrm{B}(\lambda)$, for a shell with a bumpy imperfection; \textit{i.e.}, $c=+1$ in Eq.~(\ref{eqn:gaussiandimple}).
    }
    \label{fig:fig_5}
\end{figure}

In Fig.~\ref{fig:fig_5}, to characterize the knockdown factor behavior with respect to the defect width, we present $\kappa_\mathrm{D}$ for dimpled shells (Fig.~\ref{fig:fig_5}a), and $\kappa_\mathrm{B}$ for bumpy shells (Fig.~\ref{fig:fig_5}b), as functions of $\lambda$. The results are qualitatively the same as in Fig.~\ref{fig:fig_4}. In the case of dimpled shells (Fig.~\ref{fig:fig_5}a), for small defect amplitudes, $\overline{\delta} \leq 3$, the $\kappa_\mathrm{D}(\lambda)$ curves are  non-monotonic. First, $\kappa_\mathrm{D}(\lambda)$ decreases until a threshold defect amplitude and then increases. However, for larger defect amplitudes, $\overline{\delta} \geq 3$, $\kappa_\mathrm{D}$ decreases monotonically. We highlight the fact that the threshold defect amplitude, $\overline{\delta} \approx 3$, corresponds to the largest dimple amplitude before the onset of any of the plateaux for the whole range of $\lambda$ considered. Past $\overline{\delta} \approx 3$, the $\kappa_\mathrm{D}(\lambda)$ curves are monotonic due to the insensitivity of shells to defect amplitude in this regime, for all $\lambda$ values explored (cf. Fig.~\ref{fig:fig_4}a).

Turning to bumpy defects, in Fig.~\ref{fig:fig_5}(b), we plot $\kappa_\mathrm{B}$ versus $\lambda$, noting that the behavior is different than their dimpled counterpart (Fig.~\ref{fig:fig_5}a). We find that $\kappa_\mathrm{B}(\lambda)$ is always non-monotonic, decreasing up to $\lambda \lesssim 2.5$, and then increasing for $\lambda \gtrsim 2.5 $. By contrast, for the dimpled shells (Fig.~\ref{fig:fig_5}a), $\kappa_\mathrm{D}(\lambda)$ was only non-monotonic when $\overline{\delta} \leq 3$.
This distinguishing feature between bumpy and dimpled shells can be attributed to the fact that, in the dimple case, the plateau region is  insensitive to defect amplitude, when $\overline{\delta} \geq 3$ for all values of $\lambda$; this behavior does not exist in bumpy shells given the absence of any plateauing.

\begin{figure*}[ht!]
    \centering
    \includegraphics[width=0.75\textwidth]{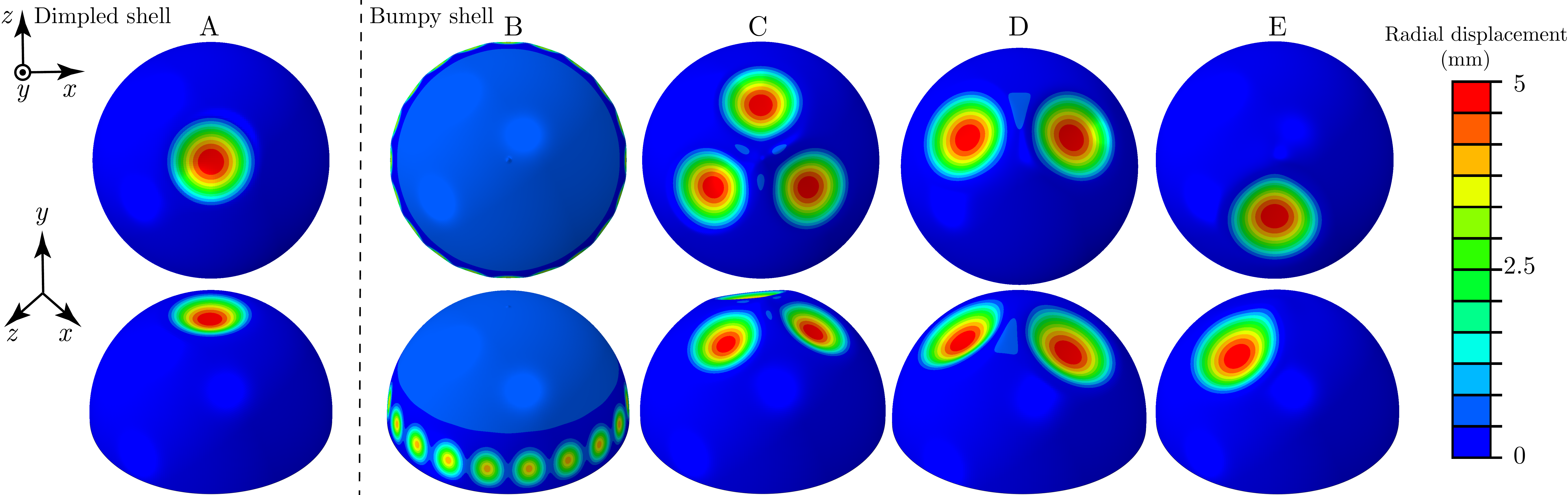}
    \caption{Representative post-buckling configurations. (A) A dimpled post-buckling configuration is representative of all shells containing a dimpled defect (even if the exact values of the radial displacement may differ). (B, C, D, E) Post-buckling configurations of shells containing a bumpy defect, for the selected cases of $\lambda=2.5$ and $\overline{\delta}=\{0.3,\,1.3,\,2.9,\,4.4\}$, respectively. 
    }
    \label{fig:fig_6}
\end{figure*}

Representative snapshots of post-buckling configurations obtained in the FEM simulations are shown in Fig.~\ref{fig:fig_6}; the color map represents radial displacements. The top ($x$-$y$) view of the shells is presented in the top row and the isometric ($x$-$y$-$z$) view is in the lower row. We refer to the post-buckling configuration as the first stable mode captured along the pressure-volume path~\cite{lee2016geometric} immediately after the onset of buckling. By way of example, in Fig.~\ref{fig:fig_6}, we consider  imperfect shells containing a dimpled defect with $\lambda=2.5$ and $\overline{\delta}=1.8$, in panel (A), and  bumpy defects with $\lambda=2.5$ and $\overline{\delta}=\{0.3,\, 1.3,\, 2.9,\, 4.4\}$, in panels (B)-(E), respectively. The axisymmetric post-buckling configuration in Fig.~\ref{fig:fig_6}(A) is representative of all the dimpled imperfect shells within the explored range of parameters: the buckling initiates at the defect location and expands axisymmetrically outwards. The post-buckling configurations are qualitatively distinct for shells with bumpy defects and depend on the value of $\overline{\delta}$; see Fig.~\ref{fig:fig_6}(B)-(E). For small defect amplitudes (\textit{e.g.}, $\overline{\delta}=0.3$, B), the shell buckles with a periodic deformation mode (akin to wrinkling) near the clamped equator, far from the bumpy defect located at the north pole. It is possible these results for small-imperfection shells are dominated by imperfections induced by the clamping conditions or by numerical imperfections (artifacts) caused by the meshing. However, in the experimental observations of Ref.~\cite{derveni2022probabilistic}, we did find that the buckling location is close to the boundary for small bumpy defects, which would tend to suggest that the periodic deformation mode is not an artifact. For higher values of $\overline{\delta}$, the loci of buckling occur near the bumpy defect but non-axisymmetrically to its side. For example, these post-buckling configurations are lobed with three, two, or one inverted-cap region for $\overline{\delta}=1.3,\,2.9,\,$ and $4.4\,$, respectively. A detailed analysis of these post-buckling configurations for bumpy shells is beyond the scope of the present study.
\begin{figure}[h!]
    \centering
    \includegraphics[width=0.9\columnwidth]{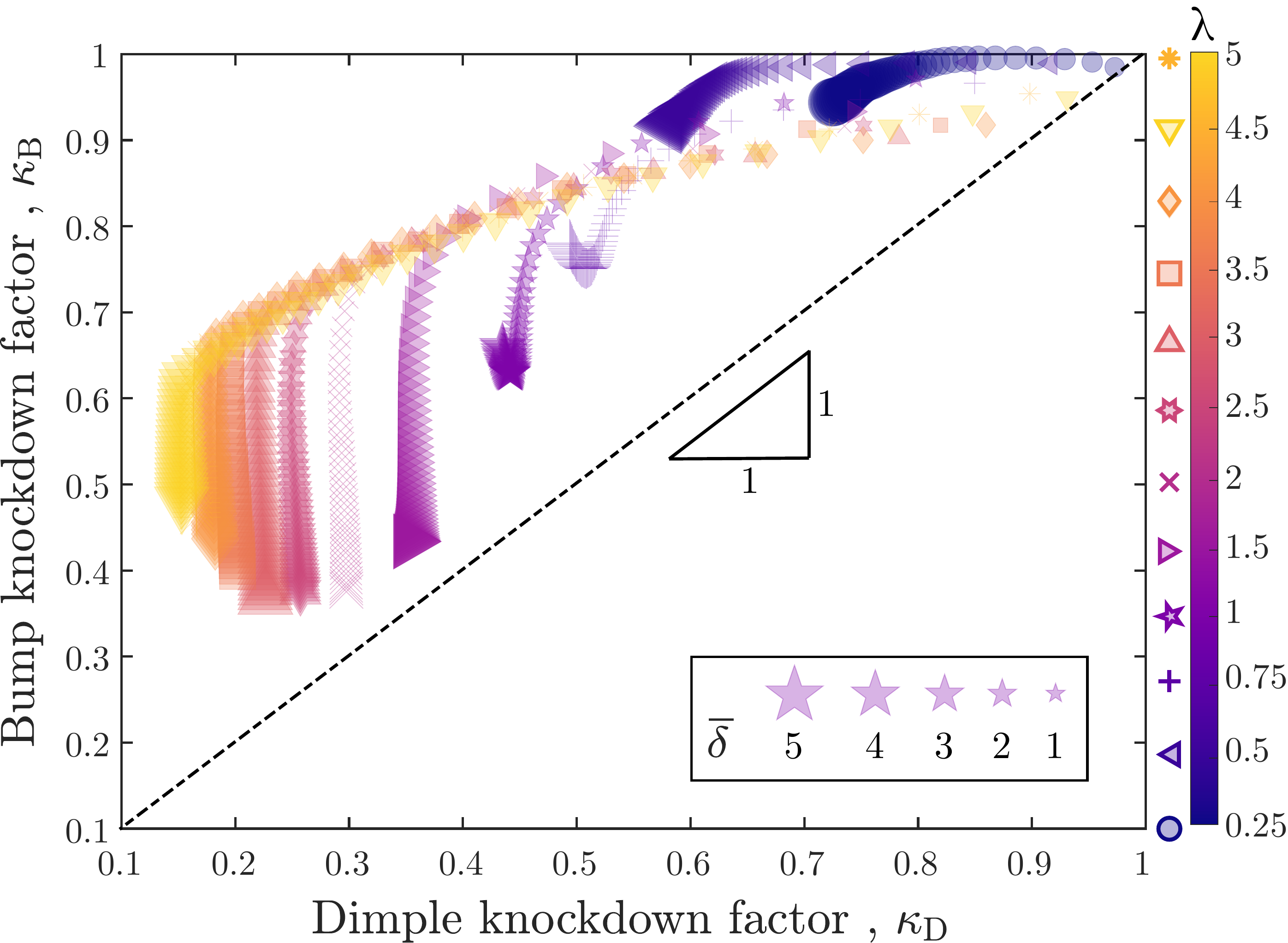}
    \caption{Knockdown factor of bumpy shells, $\kappa_\mathrm{B}$, versus that of dimpled shells, $\kappa_\mathrm{D}$, for a range of dimensionless defect geometric parameters, $0.25\leq \lambda \leq 5$, and defect amplitudes, $0.1 \leq \overline{\delta} \leq 5$. The values of $\lambda$ are color-coded (see color bar), and the values of $\overline{\delta}$ are represented by the size of the symbol (see legend). The dashed line represents $\kappa_\mathrm{B}=\kappa_\mathrm{D}$.}
    \label{fig:fig_7}
\end{figure}

Finally, for an even more direct comparison between the dimpled and bumpy cases, in Fig.~\ref{fig:fig_7}, we convey an alternative representation of the same data reported above by plotting $\kappa_\mathrm{D}$ as a function of $\kappa_\mathrm{B}$. Each data point corresponds to the same pair of $(\overline{\delta},\,\lambda)$ parameters for bumps and dimples. Different marker symbols and colors define various values of $\lambda$, while the marker size indicates the variation of $\overline{\delta}$. Beyond the specific quantitative observations uncovered from the data in Figs.~\ref{fig:fig_4} and~\ref{fig:fig_5}, this representation highlights that bumpy shells consistently have a higher buckling strength than dimpled shells, with all of the data lying above the $\kappa_\mathrm{B} = \kappa_\mathrm{D}$ line (dashed line in Fig.~\ref{fig:fig_7}). Three different regimes of behavior are observed, similarly to Fig.~\ref{fig:fig_4}(b). First, for $\lambda=\{0.25, 0.5\}$, both $\kappa_\mathrm{B}$ and $\kappa_\mathrm{D}$ decrease with increasing defect amplitude, but the reduction in $\kappa_\mathrm{D}$ is more pronounced than $\kappa_\mathrm{B}$ ought to the lower sensitivity of bumps to defect amplitude in this regime.  Second, for $\lambda=\{0.75, 1\}$, we observe a non-monotonic behavior; with increasing $\overline{\delta}$, first, both $\kappa_\mathrm{D}$ and $\kappa_\mathrm{B}$ decrease until a specific value of $\overline{\delta}$ after which, $\kappa_\mathrm{B}$ increases, while $\kappa_\mathrm{D}$ continues to decreases. Two distinct regions are obvious in the third and last regime for $\lambda \geq 1.5$. Initially, decreasing $\kappa_\mathrm{D}$ follows a decrease in $\kappa_\mathrm{B}$ until the defect amplitude of the plateau onset~\cite{jimenez_technical_2017}. After this onset, $\kappa_\mathrm{B}$ continues to decrease while $\kappa_\mathrm{D}$ remains approximately unchanged (plateau region of insensitivity to defect amplitude).
\section{\label{sec:level1}Discussion}
\label{sec:discussion}
In this section, we provide a discussion that seeks to address, even if speculatively, the following emerging questions: Why are bumps stronger than dimples? Why do bumps show different buckling modes of deformation compared to dimples?

The dimpled and bumpy shells are only distinguishable by their defect region located at the pole, with $c=\pm1$ in Eq.~(\ref{eqn:gaussiandimple}). We focus on the difference in the geometry of their undeformed (initial) configuration, as measured by the mean and Gaussian curvatures profiles defined, respectively, as 
\begin{equation}
\begin{aligned}
   & \mathcal{K}_\mathrm{H}(\beta)= \frac{1}{2}(k_1+k_2), \\
   & \mathcal{K}_\mathrm{G}(\beta)= k_1 k_2,
\end{aligned}
\end{equation}
where $k_1$ and $k_2$ are the two  principal (local) curvatures of the shell surface. We have numerically computed $\mathcal{K}_\mathrm{H}$ and $\mathcal{K}_\mathrm{G}$ with the function~\texttt{surfature}~\cite{surfatureClaxton} in~\textit{MATLAB}, taking as input the point-cloud data representation of the undeformed surface that input into the FEM simulations. We will add the subscripts D and B to denote the corresponding quantities for dimples and bumps, respectively, \textit{i.e.,} $(\mathcal{K}_\mathrm{HD},\,\mathcal{K}_\mathrm{HB})$ and $(\mathcal{K}_\mathrm{GD},\,\mathcal{K}_\mathrm{GB})$.

\begin{figure}[h!]
    \centering
    \includegraphics[width=\columnwidth]{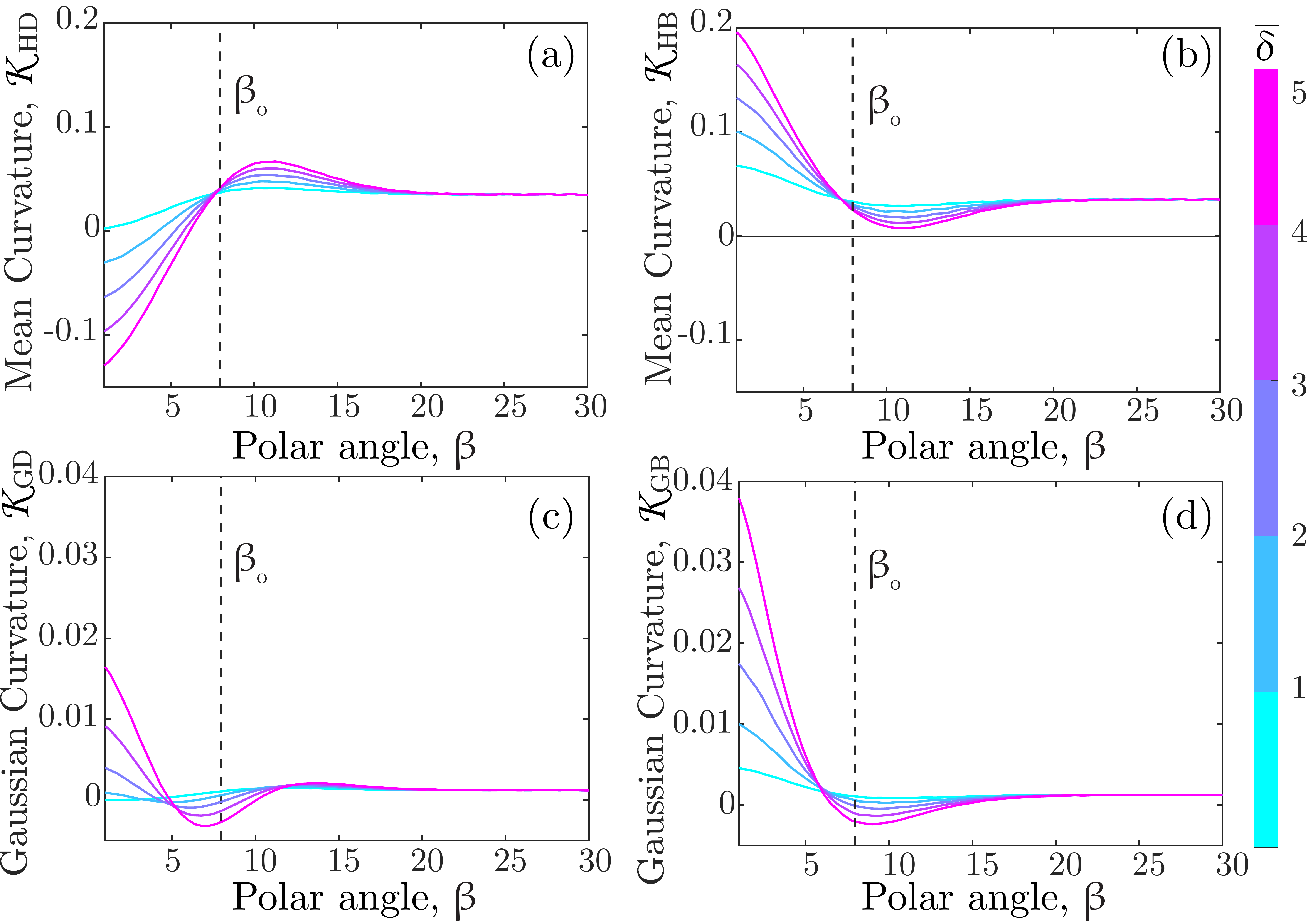}
    \caption{Curvature profiles of the initial geometry of the imperfect shell as a function of the polar angle, $\beta$. Representative cases are chosen with the dimensionless geometric parameter of $\lambda=2.5$, and amplitudes in a range of $\overline{\delta}\in[1,\,5]$ (see color bar). (a) Mean curvature, $\mathcal{K}_\mathrm{HD}$, and (c) Gauss curvature, $\mathcal{K}_\mathrm{GD}$, for a shell with a dimpled imperfection; \textit{i.e.}, $c=-1$  in Eq.~(\ref{eqn:gaussiandimple}). (b) Mean curvature, $\mathcal{K}_\mathrm{HB}$, and (d) Gauss curvature, $\mathcal{K}_\mathrm{GB}$, for a shell with a bumpy imperfection; \textit{i.e.}, $c=+1$  in Eq.~(\ref{eqn:gaussiandimple}). The vertical dashed line indicates $\beta_{\circ}$, the defect opening angle corresponding to $\lambda=2.5$.}
    \label{fig:fig_8}
\end{figure}

In Fig.~\ref{fig:fig_8}, we plot the mean curvature, $\mathcal{K}_\mathrm{H}$, in panels (a), and (b), and the Gaussian curvature, $\mathcal{K}_\mathrm{G}$, in panels (c) and (d), as functions of the polar angle, $\beta$. The angular width of the defect, $\beta_{\circ}$, defined in Eq.~(\ref{eqn:gaussiandimple}), is represented by the vertical dashed lines. We restrict our results to the representative case with $\lambda=2.5$ (where the knockdown factor of bumpy shells is lowest) while varying the defect amplitudes $\overline{\delta}=\{1,\, 2,\, 3,\, 4,\, 5\}$ (see color bar). The left panels (a) and (c) correspond to the dimpled shells, and the right panels (b) and (d) to the bumpy shells. Qualitatively similar behavior to what we describe next is found for other values of $\lambda$, but a detailed quantitative analysis is beyond the scope of the present work and unnecessary to the qualitative interpretation that we will provide. 

Hereon, we shall refer to the $\beta<\beta_{\circ}$ region as the \textit{core} of the defect and to the neighboring region right past the defect, $\beta\gtrsim\beta_{\circ}$, as the \textit{rim} of the defect.
The mean-curvature curves for dimpled shells, $\mathcal{K}_\mathrm{HD}(\beta)$ (Fig.~\ref{fig:fig_8}a), exhibit a maximum located at the defect rim.
Within the defect core, by construction, the dimples have a minimum mean curvature that is typically negative and always lower than that of the nominal spherical shell. By contrast, for the bumpy shells, all the $\mathcal{K}_\mathrm{HB}(\beta)$ curves have a minimum located at the defect rim (Fig.~\ref{fig:fig_8}b).
At the defect core, the bumps have positive mean curvature, always greater than that far away in the shell. 

Rewording the above observations, it is important to highlight that the minimum of $\mathcal{K}_\mathrm{H}$ occurs at the core for dimples and at the rim for bumps. Conversely, the maximum of $\mathcal{K}_\mathrm{H}$ occurs at the rim for dimples and at the core for bumps. As evidenced in Fig.~\ref{fig:fig_6} and studied extensively in the literature, a depressurized imperfect spherical shell exhibits a buckling mode with one (or more) inward-inverted cap, whose mean curvature has the opposite sign of the nominal sphere. It is reasonable to envision that regions of mean curvature lower (or higher) than that of the nominal sphere will serve as weak (or strong) spots, respectively. This reasoning, even if speculative, is compatible with the results in Fig.~\ref{fig:fig_6}. For dimples (Fig.~\ref{fig:fig_6}A), the post-buckling configuration does indeed occur at the defect core, where $\mathcal{K}_\mathrm{H}$ is minimum. For bumps (Fig.~\ref{fig:fig_6}B-D), the buckling appears to nucleate at the defect rim, where $\mathcal{K}_\mathrm{H}$ is minimum, and repelled by the defect core, which appears to have a stiffening effect. Moreover, the fact that $\mathcal{K}_\mathrm{H}$ is always positive in the considered range of $\overline{\delta}$  may be the source of why the knockdown factor of bumpy shells is consistently higher than that of dimpled shells.

Regarding the Gaussian curvature data presented in Fig.~\ref{fig:fig_8}(c, d), the results are, as far as we can tell, less insightful. We observe that at the defect core, $\mathcal{K}_\mathrm{G}$ is higher for the bumpy than the dimpled shells, which may further contribute to the lower buckling strength of the latter (for the same magnitude of geometric parameters). Otherwise, both cases display Gaussian curvature profiles that are qualitatively similar. All $\mathcal{K}_\mathrm{G}$ curves are non-monotonic with a minimum near the defect rim, occurring before (or after) $\beta_{\circ}$ for dimples (or bumps), respectively. In both cases, this minimum can be negative for defects with larger amplitudes ($\overline{\delta}\gtrsim 1$ for the dimples and $\overline{\delta}\gtrsim 3$ for the bumps) but always positive otherwise. Outside of this region of the minimum neighboring the rim, $\mathcal{K}_\mathrm{G}>0$ in both cases. Overall, we do not see any salient qualitative differences in the $\mathcal{K}_\mathrm{G}$ between the dimpled and bumpy cases that correlate to the $\kappa_D>\kappa_B$ reported in Fig.~\ref{fig:fig_7} and earlier plots.

There are similarities between the geometry of the bumpy shells we considered and the classical literature for the Cohn-Vossen shape~\cite{UnstarreCohn}. Shells with non-constant and sign-changing Gaussian curvature can be a source of an exceptional bending mode on a surface of revolution~\cite{audoly2010elasticity}. It is possible that this behavior can be related to the different buckling modes we observed in our bumpy shells, although we have no formal ground other than reasoning by analogy to support this statement. Future theoretical work will be necessary to further rationalize the present findings, which point to the importance of the detailed curvature profiles of doubly curved imperfect shells, with a special spotlight on their mean curvature.
\section{\label{sec:level1}Conclusion}
\label{sec:discussion}
We used an existing finite-element simulations approach, which was validated previously against experiments, to study the buckling strength of imperfect shells containing either a dimpled or a bumpy imperfection. Whereas dimpled shells have been studied previously in much detail, bumpy shells have remained largely unexplored. We considered defects with a standard Gaussian profile (cf. Eq.~\ref{eqn:gaussiandimple}), enabling direct and detailed comparisons across the dimpled ($c=-1$) and bumpy ($c=+1$) cases. Our results evidence that the role of bumps in reducing the buckling strength of the spherical shell is less dramatic than for dimples, at least within the ranges of parameters we explored. The knockdown-factor sensitivity to the detailed defect geometry is also less prominent in bumps. Overall, the knockdown factor of a bumpy shell is always greater than that of a dimpled one, $\kappa_\mathrm{B} > \kappa_\mathrm{D}$, for the same magnitude of geometric parameters. In both cases, the knockdown factor is not always reduced when the defect is widened.

We attempted to discuss the differences in knockdown factor between dimpled and bumpy shells under the light of their mean and Gaussian curvature profiles. Our interpretation suggests that regions of the imperfect shell with minimal mean curvature serve as weak points for the onset of buckling. These minima occur at the defect core for dimpled shells and at the defect rim for bumpy shells. For the latter, the core appears to have a stiffening effect, which repels the post-buckling inverted caps making the buckling mode asymmetric and potentially multi-lobed.  

We acknowledge that our investigation was mostly descriptive and observational. In the absence of formal theoretical framework, it is difficult to devise a predictive rationale for these observations. Still, we hope that our thorough comparative study will be 
valuable in the ongoing revival of shell-buckling studies. A systematic theoretical investigation will be a much-needed next step in rationalizing the current findings. 
Additionally, it would be interesting to consider other imperfection geometries and establish direct relations between the mean/Gaussian curvature profiles and the resulting critical buckling conditions. Shell buckling is a highly nonlinear and nontrivial phenomenon, and we believe that specific case studies like ours are essential to gaining insight and motivating modeling directions.

From a practical viewpoint, our research is aligned with efforts currently underway by NASA and others interested in large-scale shell structures~\cite{hilburger2006shell, hilburger2012developing, castro2014geometric}. These efforts aim to replace the purely-empirical knockdown factors guidelines in design codes of aerospace structures with mechanics-based predictive methods that take manufacturing-based data on the imperfection distributions as input. Our results demonstrate that different types of defects, even if characterized by similar geometric parameters, can yield quantitatively and qualitatively different reductions of buckling strength. For example, the design of shells containing only bumpy defects can be tackled less conservatively than dimpled shells.
\begin{acknowledgments}
A.A. thanks the Federal Commission for Scholarships for Foreign Students (FCS) support through a Swiss Government Excellence Scholarship (Grant No. 2019.0619). We are also grateful to John Hutchinson for his encouragement. During the initial stages of our investigation, he shared with us earlier preliminary findings of his suggesting a distinct buckling behavior between dimpled and bumpy shells. These nurturing interactions drove our motivation to tackle the thorough comparison in the present study. 
\end{acknowledgments}
\nocite{*}

\bibliography{apssamp}

\end{document}